\documentclass[aps,twocolumn,showpacs]{revtex4}% Physical Review B
\usepackage{graphicx}% Include figure files
\begin{document}
\title{Mean--field results on the Anderson impurity model out of equilibrium}
\author{A. Komnik$^{1}$ and A. O. Gogolin$^{2}$}
\affiliation{$^{1}$Physikalisches Institut, Albert--Ludwigs--Universit\"at,
D--79104 Freiburg, Germany \\
$^{2}$Department of Mathematics, Imperial College London, 180 Queen's Gate,
London SW7 2BZ, United Kingdom}
\date{\today}

\begin{abstract}
We investigate the mean--field phase diagram of the Anderson impurity model
out of equilibrium. Generalising the unrestricted Hartree-Fock
approach to the non-equilibrium situation we derive and analyse
the system of equations defining the critical surface separating the
magnetic regime from the non-magnetic one. An exact analytic
solution for the phase boundary as a function of the applied
voltage is found in the symmetric case. Surprisingly, 
we find that as soon as there is an asymmetry, even small,
between the contacts, no finite
voltage is able to destroy the magnetic regime which persists 
at arbitrary high voltages.
\end{abstract}
%CHECK THE PACS !!!

\maketitle
%\section{Introduction} \label{introduction}
The field of `Kondo physics' sprung
from the crucial insight by Anderson into the formation of local 
magnetic moments in metals \cite{andersonspin}. 
In Ref.\cite{andersonspin}, Anderson formulated a very simple 
model, now known as Anderson's impurity model (AIM), and
showed that there is a critical value of Coulomb coupling
($U_c$) above which a local magnetic moment spontaneously forms
(in the framework of a mean-field approximation).
It was understood later (see \cite{hewson}, especially
Chapter 5, and references therein), by using more 
sophisticated methods, that $U_c$ is in fact a 
crossover rather than a sharp `phase transition'.
Nevertheless, the Anderson's original work \cite{andersonspin}
is a milestone for our understanding of the `standard' Kondo
problem. It still serves as a foundation for
various more refined techniques, a notable example being
the `local moment approach' (LMA)\cite{logan}. 

In recent years, considerable theoretical interest 
in the non-equilibrium AIM developed, fuelled by experimental results
on non-equilibrium transport in quantum dots where
Kondo effect signatures were observed \cite{goldhaber-gordon,kouwenhoven}.
The low bias voltage ($V$) limit of the AIM has been almost
completely understood \cite{kamglazman,oguri}
(perhaps the precise mechanism of `dephasing' still being 
in need of some clarification \cite{coleman}).
When the voltage becomes larger than some characteristic scale
(like the Kondo temperature), the situation is less clear.
While there are some interesting results in the literature \cite{rosch},
no consistent theoretical method seems to be
available to tackle the problem beyond the perturbation expansion in the
on-site interaction, see e.g. Ref.\cite{ueda}. In this paper
we provide a starting point for one. Our objective
is to generalise Anderson's original analysis
[unrestricted Hartree--Fock (HF) approach] 
to the out of equilibrium model. 
Note that our starting point is complementary to that 
suggested in Ref.\cite{parcollet}, which is appropriate
for an expansion in the electron tunnelling amplitudes.

We start with the standard AIM Hamiltonian. It contains a single
site with energy $\Delta$ described by electron creation and
annihilation operators $d^\dag_\sigma$, $d_\sigma$, where $\sigma$
denotes the spin orientation. This site is coupled to two
non-interacting electron systems $i=1,2$, which we  shall
also call `leads'. Their dynamics is governed by the Hamiltonians
$H_0[\psi_{i \sigma}]$, where $\psi_{i \sigma}$ are the
corresponding electron field operators. The coupling of the leads
to the impurity site takes place via hopping with energy-independent
amplitudes $\gamma_{1,2}$,
\begin{eqnarray}                      \label{Ham0}
 H &=&  \sum_{\sigma=\uparrow,\downarrow} \Big\{ \Delta
 d^\dag_\sigma d_\sigma +  \sum_{i=1,2} \Big\{ H_0[\psi_{i \sigma}] +
 \nonumber \\
 &+& \gamma_i[
d^\dag_\sigma \psi_{i \sigma}(0) + \psi_{i \sigma}^\dag(0) d_\sigma
 ]\Big\}\Big\}
+ U d^\dag_\uparrow d_\uparrow \, d^\dag_\downarrow d_\downarrow
\, .
\end{eqnarray}
The last term describes the on-site repulsion on
the impurity. The cornerstone of the HF approach
is the following approximation for this last term,
\begin{eqnarray}\label{Hfdefinition}
 U d^\dag_\uparrow d_\uparrow d^\dag_\downarrow d_\downarrow
\rightarrow U ( n_{\uparrow} d^\dag_\downarrow d_\downarrow +
n_{\downarrow} d^\dag_\uparrow d_\uparrow)  \, ,
\end{eqnarray}
where $n_{\sigma} = \langle d^\dag_{\sigma} d_\sigma \rangle$ are
the level population probabilities which have to be found
self-consistently. The self-consistency equations
are basically the population probabilities as functions of each
other. In order to calculate them we first fix both $n_\sigma$. Then the
full Hamiltonian separates with respect to spin degrees of freedom
and can easily be diagonalised keeping in mind
that $n_\sigma$ are parameters. Since we are dealing with a
non-equilibrium situation, the calculation of the
population probabilities can only be accomplished using suitable 
modifications of the conventional equilibrium techniques. We decided to use 
the Keldysh 
formalism \cite{keldysh,LLX}. Let us first define the local (taken
at the site adjacent to the impurity) Green's function of the
lead electrons (from now on we set the spacial coordinate to $x=0$ and 
drop it hereafter),
\begin{eqnarray}             \label{genKeldysh}
 G_{i \sigma}(t-t') = - i \langle T_C \psi_{i \sigma}(t) \psi_{i
 \sigma}^\dag(t') \rangle \, ,
\end{eqnarray}
where $T_C$ denotes the time ordering operation along the closed Keldysh
contour $C$ consisting of the forward path $C_-$, running from $-\infty$ to
$\infty$, and the backward path $C_+$ going the other way round. After an
appropriate placement of the time variables, one can recover all
usual Keldysh Green's functions from the
generalised function (\ref{genKeldysh}), we use notation of Ref.\cite{LLX}. 
We shall use only the bare Green's functions of the
lead electrons, i.~e. those calculated without the tunnelling
$\gamma_i=0$ [the upper(lower) sign and subscript 1(2) corresponds to the left
(right) lead],
 \begin{eqnarray}                       \label{gpsinonint}
 G^{--}_{1(2) \sigma}(\omega) &=& G^{++}_{1(2) \sigma}(\omega)= - i \, \frac{\rho_0}{2}
 \, \mbox{sgn}(\omega \pm V/2) \, , \nonumber \\
 G^{-+}_{1(2) \sigma}(\omega) &=& i \, \rho_0 \, \Theta(-\omega \mp V/2 ) \, , \nonumber \\
 G^{+-}_{1(2) \sigma}(\omega) &=& -i \, \rho_0 \, \Theta(\omega \pm V/2) \, ,
\end{eqnarray}
where $\rho_0$ is the constant density of states in the leads and
$\Theta$ denotes the Heaviside step function.

\begin{figure}[]
\vspace*{0.8cm}
\includegraphics[scale=0.3]{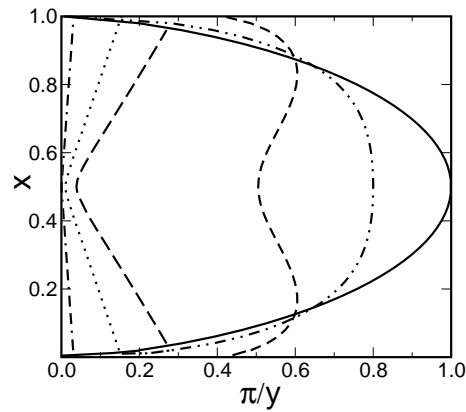}
\caption[]{\label{crosssections} Mean--field phase diagram of the AIM in the
symmetric case $\delta =1/2$ evaluated numerically using
Eqs.(\ref{criticalsurface}) for different voltage parameters $z$.
From right to left along the $x_c=1/2$ line: $z=0$, $0.5$, $1$,
$5$, $10$, $50$. The magnetic phase is bounded by the curves and
the axis of ordinates.}
\end{figure}

Green's functions of the impurity level are defined analogously and
are denoted by $D$. The function
$D^{-+}_\sigma(t-t')$ for $t-t' \rightarrow 0^-$, where $0^-$ is a
negative infinitesimal, is related to the population of
the level with spin projection $\sigma$,
\begin{eqnarray}                          \label{howtofindn}
 n_\sigma = \langle d^\dag_\sigma (t) d_\sigma(t) \rangle =
-i D^{-+}_\sigma(t-t')|_{t-t'=0^-} \, .
\end{eqnarray}
Therefore our goal is to calculate that particular Green's
function. As already mentioned, the Hamiltonian in the HF
approximation is quadratic, so that $D^{-+}_\sigma(t-t')$ can be
calculated either using the re-summation technique of
Ref.\cite{caroli}, or the non-equilibrium coherent state
functional integrals \cite{kamenev,ourPRL}. We choose the
latter method because of its relative simplicity. The idea is to
define a non-equilibrium counterpart of the partition function as
a functional integral over all fields entering the Hamiltonian and
integrate out all degrees of freedom apart of those corresponding
to $d_\sigma$ operators. The resulting expression will
be quadratic in $d_\sigma$'s which enables all correlation functions to be
simply read off. Let us define the following coherent state
functional integral corresponding to a Keldysh partition function
of the subsystem with spin orientation $\sigma$,
\begin{eqnarray}                         \label{bloedesZ}
 {\cal Z_\sigma} &=& \int {\cal D}d_\sigma^* {\cal D} d_\sigma
 \prod_i {\cal D}\psi_{i \sigma}^* {\cal D}
 \psi_{i \sigma} \exp \Big\{ -i \sum_i S_i
 \nonumber \\
 &-& i \int_C dt \, \Big[ d_\sigma^*(t) (i
 \partial_t + \Delta + U n_{-\sigma}) d_\sigma(t) \nonumber \\
  &+& \sum_i \gamma_i[
d^*_\sigma(t) \psi_{i \sigma}(t) + \psi_{i \sigma}^*(t)
d_\sigma(t)
 ]  \Big] \Big\} \, ,
\end{eqnarray}
where by abuse of terminology $d_\sigma^*$, $d_\sigma$ and
$\psi^*_{i \sigma}$, $\psi_{i \sigma}$ stand for the coherent
state variables of the corresponding operators. $S_i$ are the
actions of the electrons in each lead. Notice that all time
integrations are to be performed along the Keldysh contour
$C$. Since the full action is Gaussian in the $\psi_{i \sigma}$ 
fields we can integrate them out thereby generating a non-local action
for the impurity electrons.
Separating time integrations along the $C_\pm$ branches we obtain 
the following expression,
\begin{eqnarray}                          \label{Zagain}
   {\cal Z_\sigma} = \int {\cal D}\vec{d}^*_\sigma {\cal D} \vec{d}_\sigma \exp \Big\{- i \int
 d t d t' \nonumber \\ \times \vec{d}_\sigma^*(t) {\bf D}_\sigma^{-1} (t-t') \vec{d}_\sigma (t') \Big\} \, ,
\end{eqnarray}
with $\vec{d}_\sigma$ denoting a vector of $d_\sigma(t)$ for the
time variable $t$ on $C_\mp$. ${\bf D}_\sigma(t-t')$ is the
quantity we are looking for -- it is a matrix components of which
give the Keldysh Green's functions of the impurity level
operators. 
%Its inverse can directly be read off Eq.(\ref{Zagain}). 
After the Fourier transformation its inverse is given by
\begin{widetext}
\begin{eqnarray}                \label{matrixN1}
  {\bf D}_\sigma^{-1}(\omega) = \left(  \begin{array}{cc}
                  \omega + \Delta + U n_{-\sigma}+
                  \sum_i \gamma_i^2 G^{--}_{i \sigma}(\omega)  & \sum_i \gamma_i^2
                  G^{-+}_{i \sigma}(\omega)
                  \\ \\
                  \sum_i \gamma_i^2
                  G^{+-}_{i \sigma}(\omega) & - \omega - \Delta - U n_{- \sigma} +
                  \sum_i \gamma_i^2 G^{++}_{i \sigma}(\omega)
                  \end{array} \right)
\end{eqnarray}
\end{widetext}
Therefore the Green's function that we are interested in is given
by
\begin{eqnarray}                \label{usetheseDs}
 D^{-+}_\sigma (\omega) &=& -|{\bf D_\sigma (\omega)}|^{-1} \sum_i \gamma_i^2
 G^{-+}_{i \sigma}(\omega) \, .
\end{eqnarray}
$|{\bf D_\sigma(\omega)}|$ denotes the determinant of the matrix 
(\ref{matrixN1}). Inserting Eqs.(\ref{gpsinonint}) into
(\ref{usetheseDs}) yields
\begin{eqnarray}                        \label{finalD}
 D^{-+}_\sigma (\omega) = i \frac{ \Gamma_1 \Theta(-\omega+V/2) +
 \Gamma_2 \Theta(-\omega-V/2)}{(\omega + \Delta + U n_{-\sigma})^2
 + (\Gamma_1 + \Gamma_2)^2} \, ,
\end{eqnarray}
where $\Gamma_i = \pi \rho_0 \gamma_i^2$. Finally, according to
Eq.~(\ref{howtofindn}), $n_\sigma$ is just the energy integral of
(\ref{finalD}) with an appropriate pre-factor,
\begin{eqnarray}                    \label{selfconsE}
 n_{\sigma} &=& \frac{\delta}{\pi}\cot^{-1}\left( \frac{\Delta - V/2 + U n_{
 -\sigma}}{\Gamma} \right)
 \nonumber \\
 &+& \frac{1-\delta}{\pi}\cot^{-1}\left(
 \frac{\Delta + V/2 + U n_{-\sigma}}{\Gamma} \right)  \, ,
\end{eqnarray}
where $\Gamma = \Gamma_1+\Gamma_2$ and $\delta = \Gamma_1/\Gamma$
is the asymmetry parameter. Eqs.(\ref{selfconsE}) are the
self-consistency equations for the impurity level population
probabilities. In the equilibrium case 
these equations reduce to Eqs.(27) of Ref.\cite{andersonspin}.
Introducing dimensionless parameters $x=-\Delta/U$, $y=U/\Gamma$
and $z=V/2 \Gamma$ the above equations simplify to
\begin{eqnarray}                    \label{selfcon}
 n_{\sigma} &=& \frac{\delta}{\pi} \cot^{-1} [y(n_{-\sigma}-x)-z]
 \nonumber \\ &+&
 \frac{1-\delta}{\pi} \cot^{-1} [y(n_{-\sigma}-x)+z] \, .
\end{eqnarray}
It is reasonable to expect that both the magnetic ($n_{\sigma} \neq
n_{-\sigma}$) and the non-magnetic ($n_{\sigma} = n_{-\sigma}$) regimes
still exist for small $z$. The
critical ($x$--$y$) curve, however, acquires an additional 
dimension ($z$) and thus becomes a critical surface. 
Our goal is to determine its shape. 
Proceeding along the lines of Ref.\cite{andersonspin},
we derive the equations for the surface. The first
equation is essentially given by Eq.(\ref{selfcon}) for
$n_{\sigma}=n_{-\sigma}=n_c$ and the second one is just its
derivative with respect to $n_c$,
\begin{widetext}
\begin{eqnarray}                     \label{criticalsurface}
 \left\{  \begin{array}{l}
                  \pi n_c = \delta \cot^{-1} [y_c (n_c - x_c)-z] +
                  (1-\delta) \cot^{-1} [ y_c (n_c - x_c) + z)] \, ,
                  \\ \\
                  \pi/y_c = \delta \{1+[y_c(n_c-x_c)-z]^2\}^{-1} +
                  (1-\delta) \{1+[y_c(n_c-x_c)+z]^2\}^{-1} ,
                  \end{array} \right.
\end{eqnarray}
\end{widetext}
where the subscript `$c$' stands for the parameter of the critical
surface. Obviously, in the equilibrium situation ($z=0$) the
asymmetry does not matter and one obtains the result of Ref.
\cite{andersonspin},
\begin{eqnarray}                     \label{CSE}
 \left\{  \begin{array}{l}
         x_c = n_c - \sin(2 \pi n_c)/2 \pi \, ,
         \\
         \pi/y_c = \sin^2(\pi n_c),
         \end{array} \right.
\end{eqnarray}
where $n_c$ plays the role of a parametrisation and runs from
$0$ to $1$, see Fig.\ref{crosssections}. Out of equilibrium, the
equations (\ref{criticalsurface}) cannot be
reduced in the same way and need to be solved numerically
in the general case.  Yet it is possible to investigate some 
interesting special cases analytically.

Let us first concentrate on the symmetric case $\delta = 1/2$. 
Then Eqs.(\ref{criticalsurface}) are obviously symmetric with
respect to transformation $z \rightarrow -z$, so that we can
restrict ourselves to positive $z$. Next we observe that, just as
in equilibrium, there is an additional symmetry with respect
to the axis $x=1/2$: $x_c \leftrightarrow 1-x_c$ (which, of
course, corresponds to $n_c \leftrightarrow 1-n_c$). Moreover, at the
special point $x_c=1/2$ Eqs.(\ref{criticalsurface}) can be
solved exactly. The first of them fixes $n_c$ to $1/2$ but
does not impose any restrictions on $z$, so that from the
second equation we obtain,
\begin{eqnarray}                   \label{exactsolution}
 y_c = \pi (1+z^2) \, .
\end{eqnarray}
The main feature of this result is that, at least on the line
$x_c=1/2$, the magnetic phase is never completely destroyed by the
applied voltage. Another remarkable property of the phase diagram
which can be extracted from the result (\ref{exactsolution}) is
the fact that although the point $x_c=1/2$ on the equilibrium
curve corresponds to the lowest critical interaction strength $U$
and therefore to the most favorable condition for the magnetic
phase to emerge, it does not possess that property above 
the critical value $z^* = 1/\sqrt{3}$. This can be seen from the lowest order
expansion of Eq. (\ref{exactsolution}) in powers of $x_c-1/2$,
\begin{eqnarray}
 y_c \approx \pi (1+z^2)\left[ 1 + \frac{\pi^2}{4} (1-3z^2)(x_c-1/2)^2
 %+ o[(x_c-1/2)^2]
 \right]\, ,
\end{eqnarray}
which means that the phase diagram develops a dip toward the $x$
axis at $z > z^*$, see Fig.\ref{crosssections}.

Let us now analyse the behaviour of the critical curves for small
$x_c$. In that limit $n_c$ happens to be small as well so that we
can regard it as a good expansion parameter. From the first of
Eqs.(\ref{criticalsurface}) one can easily see that in such a
situation the $\cot^{-1}$ functions are vanishing so that it can
be solved after an asymptotic expansion of trigonometric
functions. The result is $y_c(n_c-x_c) = (\pi n_c)^{-1} +
\pi n_c z^2$. It can be plugged into the second equation and the
whole procedure results in the following parametric
representation,
\begin{eqnarray}
 \left\{  \begin{array}{l}
                  x_c = n_c - \left[ (\pi^2 n_c)^{-1} + n_c z^2 \right](\pi/y_c) \, ,
                  \\ \\
                  \pi/y_c = \sum\limits_{p=\pm} \{1+[(\pi n_c)^{-1} + \pi n_c z^2 + p z]^2\}^{-1}
                  \, .
                  \end{array} \right.
\end{eqnarray}
Being plotted numerically 
this curve starts from the origin and never crosses the
$x_c=0$ axis again unless $z > 0.707 \approx 1/\sqrt{2}$.
Therefore in this latter situation the finite voltage encourages
the system to become magnetic. This fact also reveals
itself as a `squeezing' of the magnetic region toward the lines
$x_c=0,1$, see Fig.\ref{crosssections}. 

To summarise the effects of the voltage in the symmetric case ($\delta=1$),
the equilibrium conditions for the stability of the magnetic phase
remain qualitatively intact as long as $z<z^*$. For larger voltages,
$z>z^*$, it becomes favourable to measure the dot level $\Delta$ 
from either of the two shifted chemical potentials $\mu_{R,L}$ in
order to preserve the local moment. Hence the dip in the phase diagram.
Yet, when $V$ is sent to infinity, the magnetic phase completely
disappears in the symmetric case. Surpisingly, the magnetic
phase persists up to $V=\infty$ in the asymmetric case ($\delta\neq 1$),
which we investigate next. 

%The change in the conditions for the stability of the magnetic phase
%can be best understood in the limit of high applied voltage, $z\gg 1$. 
%In that situation the smallest $U$ is needed to destroy the local moment
%when the energetic distance of the dot levels (the one with energy $\Delta$
%and that with energy $\Delta+U$) to the chemical potential of either of
%the leads, $\mu_{R,L}$, is the same, i. e. when $\mu_{R,L} = \Delta+U/2$.
%Obviously, this condition reproduces the one in the equilibrium, when it
%is given by $x_c=1/2$. However, in the intermediate voltage regimes a complex
%crossover behaviour emerges.

\begin{figure}[]
\vspace*{0.8cm}
\includegraphics[scale=0.35]{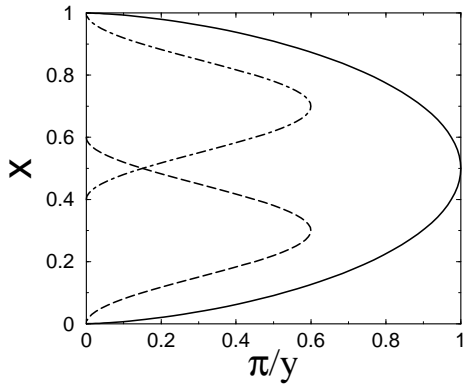}
\caption[]{\label{SymASymPlots} Mean--field phase diagram of the AIM in the
asymmetric case $\delta=0.4$ evaluated 
numerically using Eqs.(\ref{largez}). Dashed line corresponds to
high negative voltages and dot-dashed one represents the limit of high
positive voltages. For comparison: the solid line is the equilibrium $z=0$
critical curve.}
\end{figure}

The limit of large applied voltage can also be investigated
analytically. However, from the physical point of view it is more
convenient to interpret the results for a system where the voltage
is applied asymmetrically. Such situation takes place e.g. if we
shift our voltage variable to zero in one channel and to $2z$ in the
other one. Sending the voltage to infinity simplifies the first
equation of the system (\ref{criticalsurface}) to
\begin{eqnarray}                    \label{komisch}
 y_c(n_c - x_c) = \cot \left(\pi \frac{n_c-\delta}{1-\delta}\right) \, ,
\end{eqnarray}
from which we immediately see that $\delta < n_c < 1$. Insertion
of Eq. (\ref{komisch}) into the second one of the system
(\ref{criticalsurface}) yields the following parametric
representation,
\begin{eqnarray}                  \label{largez}
 \left\{  \begin{array}{l}
         x_c = n_c - (1-\delta) \sin[2 \pi (n_c-\delta)/(1-\delta)]/2 \pi \, ,
         \\ \\
         \pi/y_c = (1-\delta) \sin^2[\pi (n_c-\delta)/(1-\delta)] \, .
         \end{array} \right. 
\end{eqnarray}
At large negative voltages the picture is changed only by the
transformation $x_c \rightarrow 1-x_c$. Sending the applied
voltage to infinity effectively corresponds to a removal the weaker
lead. As one would expect in his case the critical curve has
then exactly the same shape as the curve in equilibrium apart of a
renormalisation by the asymmetry parameter $\delta$ and a shift of
the dot population $n_c$, see Fig.\ref{SymASymPlots}. 

To conclude, we investigated the mean--field phase diagram of the
non-equilibrium Anderson impurity model by means of the
unrestricted Hartree-Fock approximation. We present an 
analytic expression for the critical curve between the magnetic
and non-magnetic phases in the symmetric case when the level
energy $-\Delta$ is equal to the half of the on-site interaction
constant $U$. It turns out that above some critical value of the
applied voltage $V > 2 \Gamma /\sqrt{3}$ the critical curve
acquires a dip which means that the most favorable conditions for
the magnetic phase to form change. We show that the
magnetic phase exists at \emph{arbitrary} voltages
in the \emph{asymmetric} case. 
We consider our HF solution as an appropriate starting point
for further investigations into the problem including
fluctuations in the functional integral approach or through a 
suitable generalisation of the LMA.

%\acknowledgements
The authors would like to thank H. Grabert for an interesting 
discussion.
This work was supported by the EPSRC of the UK under grant
GR/R70309 as well as by the Landesstiftung 
Baden--W\"urttemberg gGmbH
(Germany). The authors participate in the EU network DIENOW.

\end{document}